\documentclass[aps,
pra,
reprint,
showpacs, showkeys
]{revtex4-1}
\pdfoutput=1

\usepackage{natbib}
\setcitestyle{numbers}
\usepackage{graphicx}
\usepackage{xcolor}
\usepackage{amssymb, amsmath}
\usepackage{epstopdf}
\begin{document}


\title{Cesium bright matter-wave solitons and soliton trains}


\author{Tadej Me\v{z}nar\v{s}i\v{c}}
\email[]{tadej.meznarsic@ijs.si}
\author{Tina Arh}
\author{Jure Brence}
\author{Jaka Pi\v{s}ljar}
\author{Katja Gosar}
\author{\v{Z}iga Gosar}
\author{Rok \v{Z}itko}
\author{Erik Zupani\v{c}}
\author{Peter Jegli\v{c}}\email[]{peter.jeglic@ijs.si}
\affiliation{Jo\v{z}ef Stefan Institute, Jamova 39, SI-1000 Ljubljana, Slovenia}


\date{\today}

\begin{abstract}
A study of bright matter-wave solitons of a cesium Bose-Einstein condensate (BEC) is presented.
Production of a single soliton is demonstrated and dependence of soliton atom number on the interatomic interaction is investigated. 
Formation of soliton trains in the quasi one-dimensional confinement is shown.
Additionally, fragmentation of a BEC has been observed outside confinement, in free space.
In the end a double BEC production setup for studying soliton collisions is described.

\end{abstract}


\maketitle

\section{Introduction}
\label{Sec:0:Intro}

Non-dispersing wavepackets called solitons appear in many non-linear physical systems. 
Examples of solitons can be found in water waves \cite{Russel1844}, acoustic waves \cite{Peronne2017}, light propagating through non-linear materials \cite{Chen2012}, plasmas \cite{Kourakis2005}, energy propagation along proteins \cite{Scott1992}, and many other systems including Bose-Einstein condensates (BECs) of cold atoms.
Experimental research on solitons in BECs began with creation of a dark soliton \cite{Burger1999, Denschlag2000}, followed by a bright soliton \cite{Khaykovich2002} and bright soliton trains \cite{strecker2002formation}.
Observation of more exotic gap solitons \cite{Eiermann2004}, decay of dark solitons into vortex rings \cite{Anderson2001}, interactions between solitons \cite{Weller2008, Becker2008, nguyen2014collisions}, their interactions with impurities \cite{Aycock2017}, optical potential barriers \cite{Marchant2016}, speckle potentials \cite{Boisse2017} and demonstration of a matter-wave interferometer \cite{McDonald2014} show that a cold-atom BEC is an excellent and versatile system for studying solitons.

Formation of solitons in a BEC depends on the two-body interaction between the atoms and the geometry of the trap used to confine the BEC.
A quasi-one-dimensional (quasi-1D) confinement is needed, which can be achieved in either magnetic or optical dipole traps.
In such traps a dark soliton forms as a trough of lower density within a BEC with repulsive interatomic interaction while a bright soliton is a wavepacket comprising the whole BEC with attractive interatomic interaction that can move over macroscopic distances in a vacuum.
So-called dark-bright solitons can be supported in two-component BECs, where atoms with one spin component fill the dark soliton within the BEC of the other spin component \cite{Becker2008, Dum1998, Busch2001}. 

Usually, only unchanging waves in one-dimensional integrable systems are called solitons. 
In quasi-1D harmonically confined geometry integrability is broken, but only slightly so.
The solitary waves that form from BECs are three-dimensional objects, not one-dimensional, but their propagation is limited to one-dimension.
The name soliton in this paper is used in its broader meaning comprising all localized wave packets with dispersion compensated by nonlinearity.

The fact that a bright soliton is a wavepacket that can propagate over large distances and long times without dispersion makes solitons prime candidates for use in atom interferometry. 
They have already been shown to improve the performance of Mach-Zehnder interferometer compared to regular BECs \cite{McDonald2014}.
As such a use for matter-wave solitons may be found in precision atomic sensors, e. g. gravimeters \cite{Angelis2008}, rotation sensors \cite{Barrett2014, Helm2015} and other accelerometers. 

Previously, bright solitons have been created in $^{7}$Li \cite{strecker2002formation, Khaykovich2002, Medley2014}, $^{85}$Rb \cite{Cornish2006, Marchant2013, McDonald2014} and $^{39}$K \cite{Lepoutre2016}, but this is, to our knowledge, the first demonstration of $^{133}$Cs solitons.

A study of bright $^{133}$Cs solitons and soliton trains is presented.
Our production of both is described, with the study of soliton atom number for the trains comprising of different numbers of solitons for different interaction strengths. 
Fragmentation of a BEC outside confinement is shown, with a double BEC and soliton production outlining future studies of soliton collisions.

\section{Production of Cesium Solitons}
\label{Sec:1:Solitons}

\subsection{Single soliton}
\label{Subsec:1.1:single_sol}

We produce a cesium BEC using a procedure described in detail in App.~\ref{App:PathToBEC}. 
Briefly, after the magneto-optical trap and degenerate Raman sideband cooling, we transfer the atoms into a  crossed optical dipole trap (reservoir). 
After 0.5~s of plain evaporation in the dipole trap we ramp up a tighter dimple trap, created by crossing two beams of smaller radii. 
We hold the atoms levitated with a magnetic field gradient in both traps, then turn off the dipole trap and begin with forced evaporation in the dimple trap. 
We evaporate by simultaneously decreasing the dimple beams' intensity and the magnetic field gradient until we reach the Bose-Einstein condensation at a critical temperature of around 20~nK.
We typically obtain a BEC of 5000-10000 atoms depending on the exact evaporation parameters.

\begin{figure}[t]
\centering
\includegraphics[width=1\linewidth]{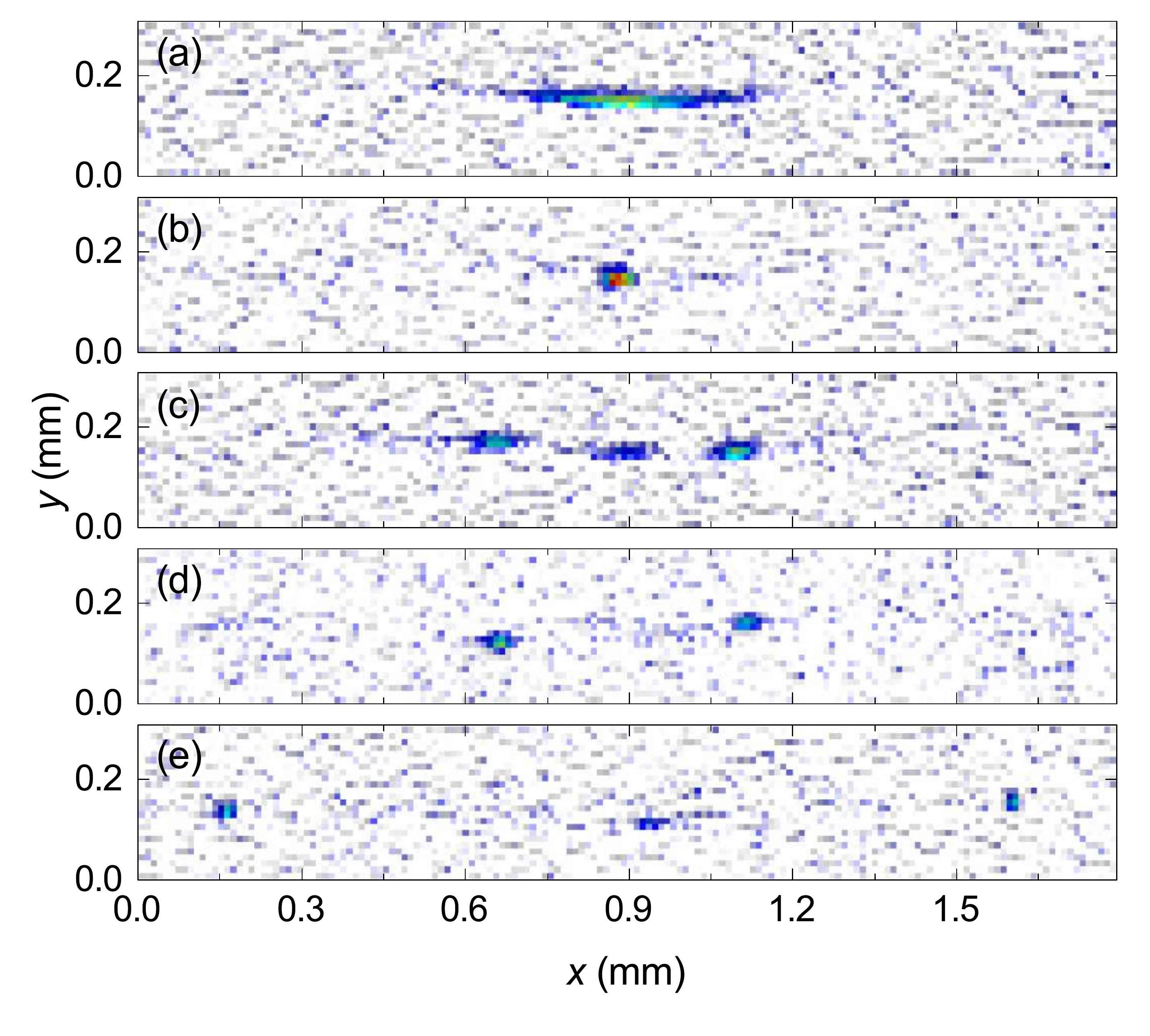}
\includegraphics[trim=0 10 0 50, clip=true, width=1\linewidth]{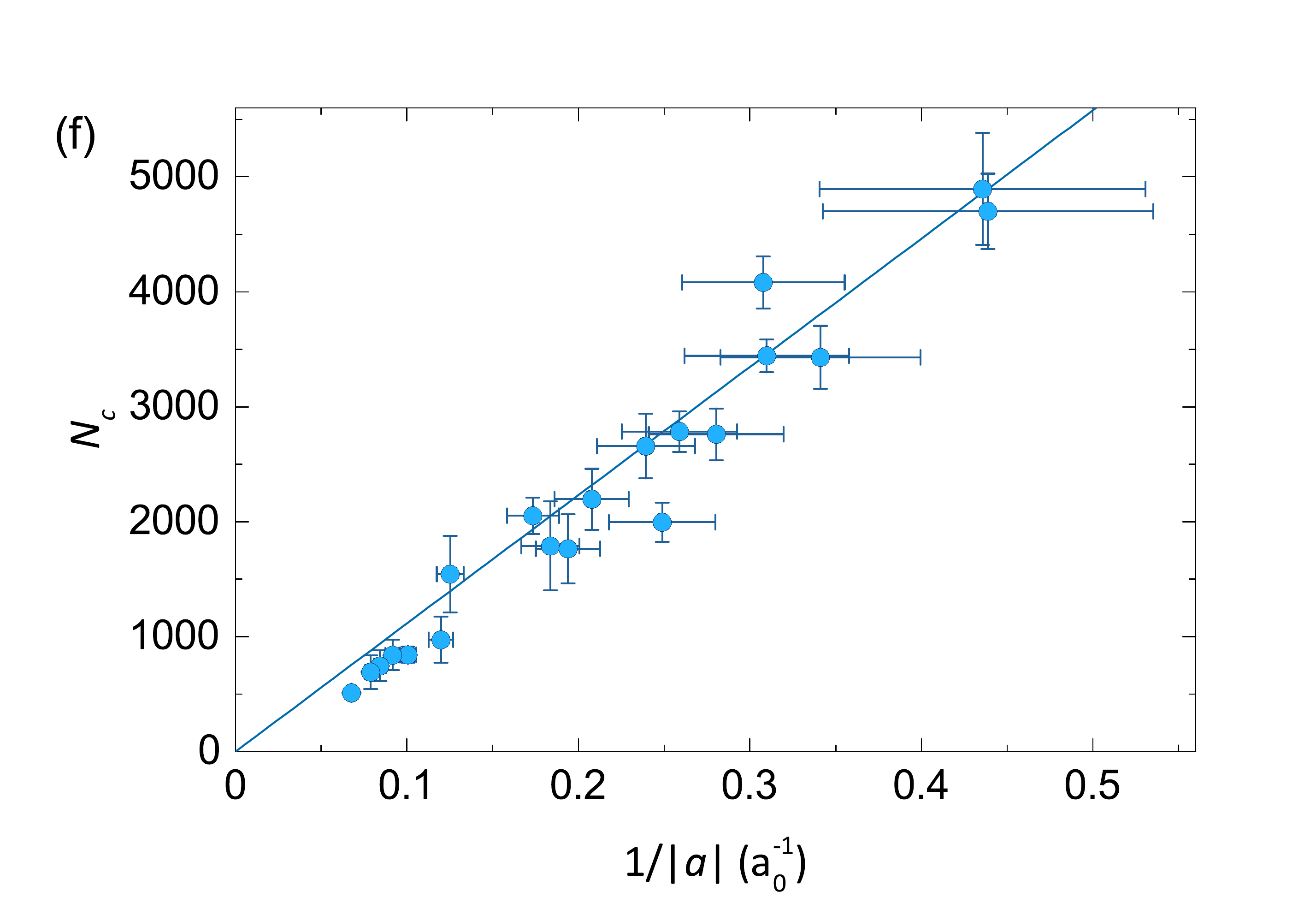}
\caption{\label{fig:1:Soliton_interaction} (a)-(e) Atoms in a quasi-1D channel 160 ms (+ 15 ms time-of-flight) after the release from the crossed dimple trap for different interatomic interactions (scattering lengths from (a)-(e): $-5.6a_0$, $-6.4a_0$, $-8.8a_0$, $-12.0a_0$, $-18.4a_0$). 
The BEC goes from dispersed in Fig.~(a), to a single soliton in Fig.~(b) and soliton trains in Figs.~(c)-(e) as the interaction becomes more and more attractive.
(f) Dependence of soliton atom number on the inverse scattering length. 
The line follows $N_c = k_c a_r/|a|$, with $k_c=0.67$ and $a_r$ determined independently of this measurement.
Vertical errorbars result from multiple measurements of $N_c$ and horizontal errorbars from the determination of the scattering length.
As the scattering length decreases, the relative error increases, which shows in the inverse scale as an increased error for larger values.
}
\end{figure}

\begin{figure}[t]
\centering
\includegraphics[width=1\linewidth]{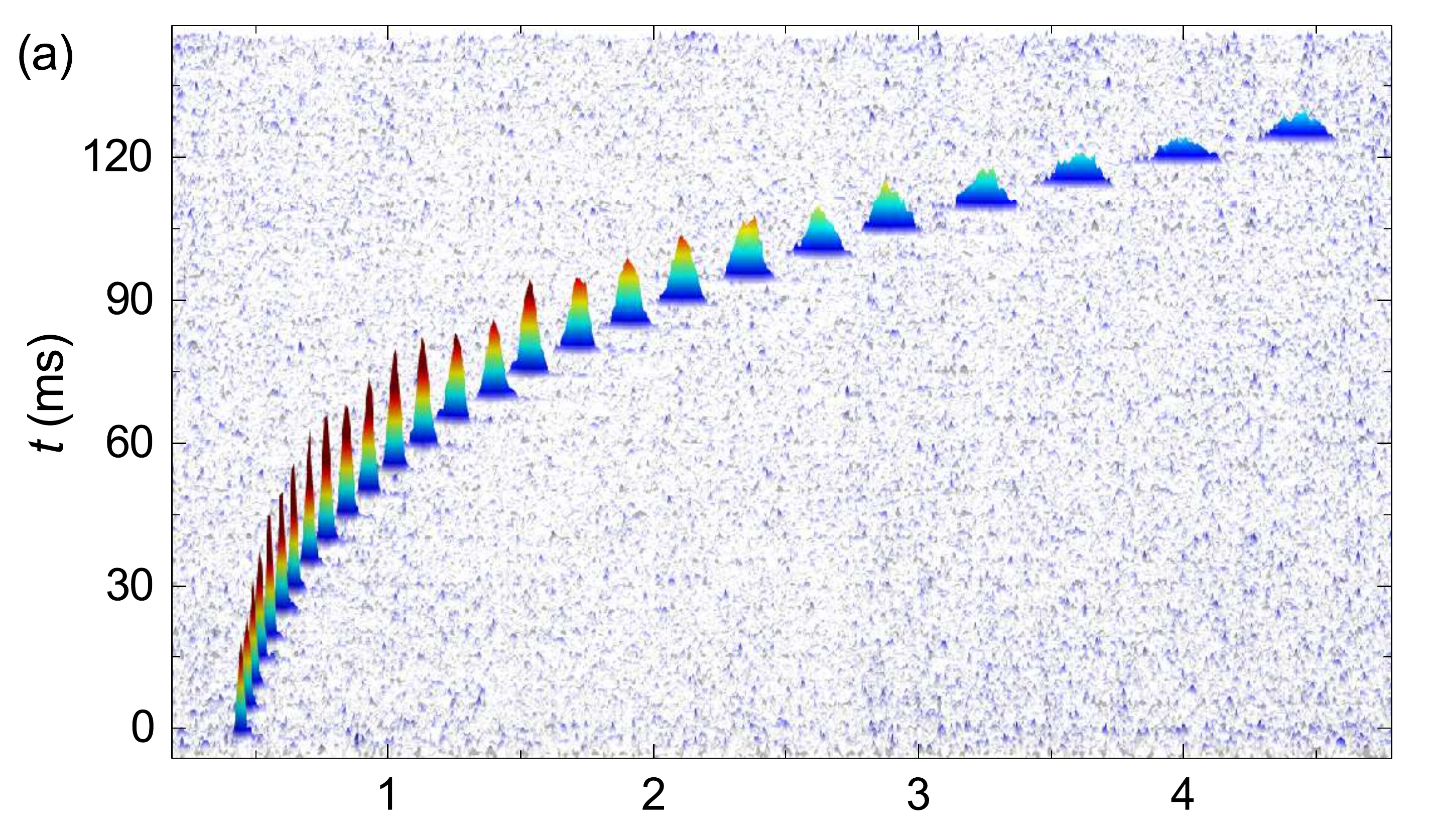}
\includegraphics[width=1\linewidth]{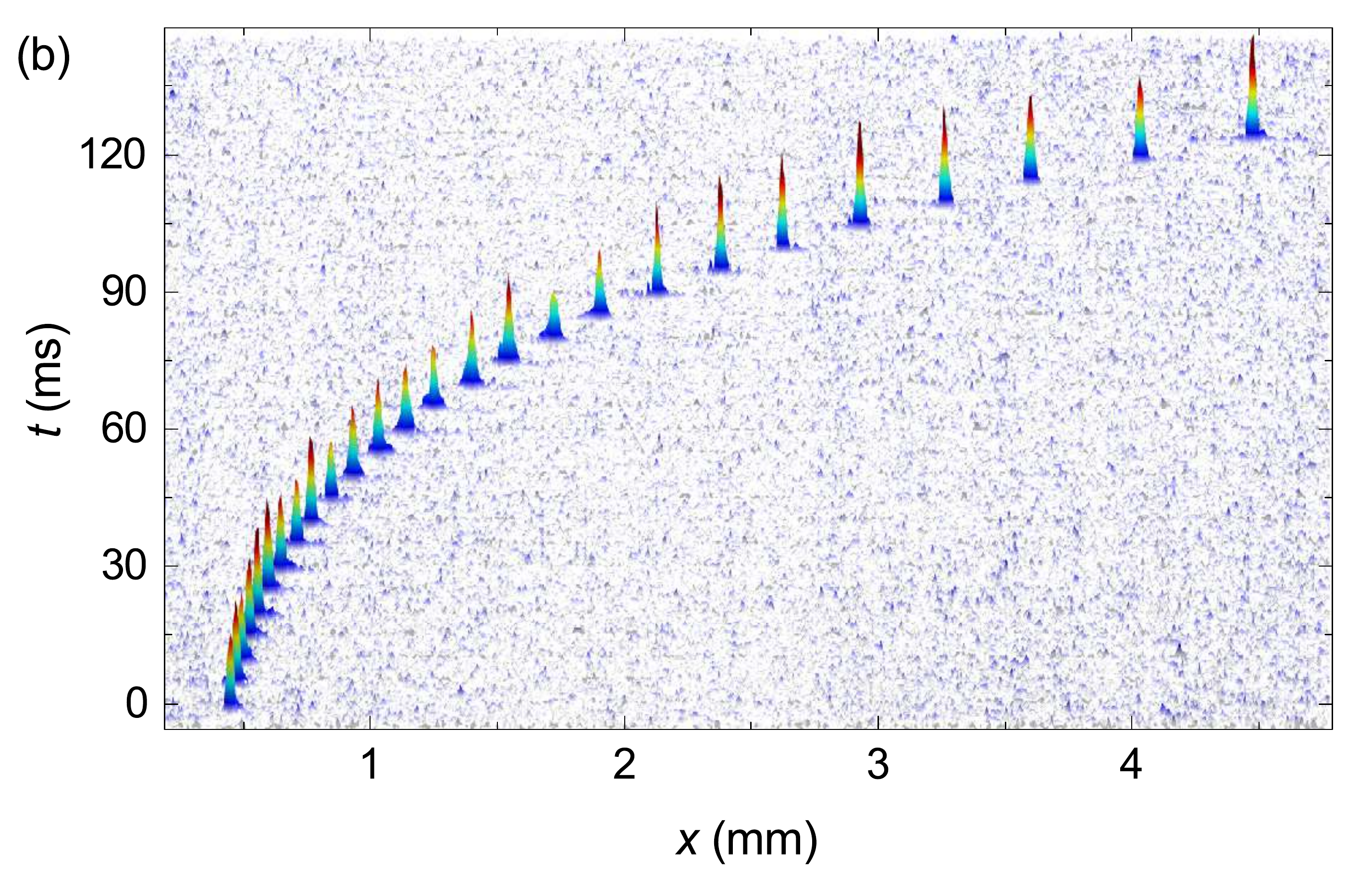}
\includegraphics[width=1\linewidth]{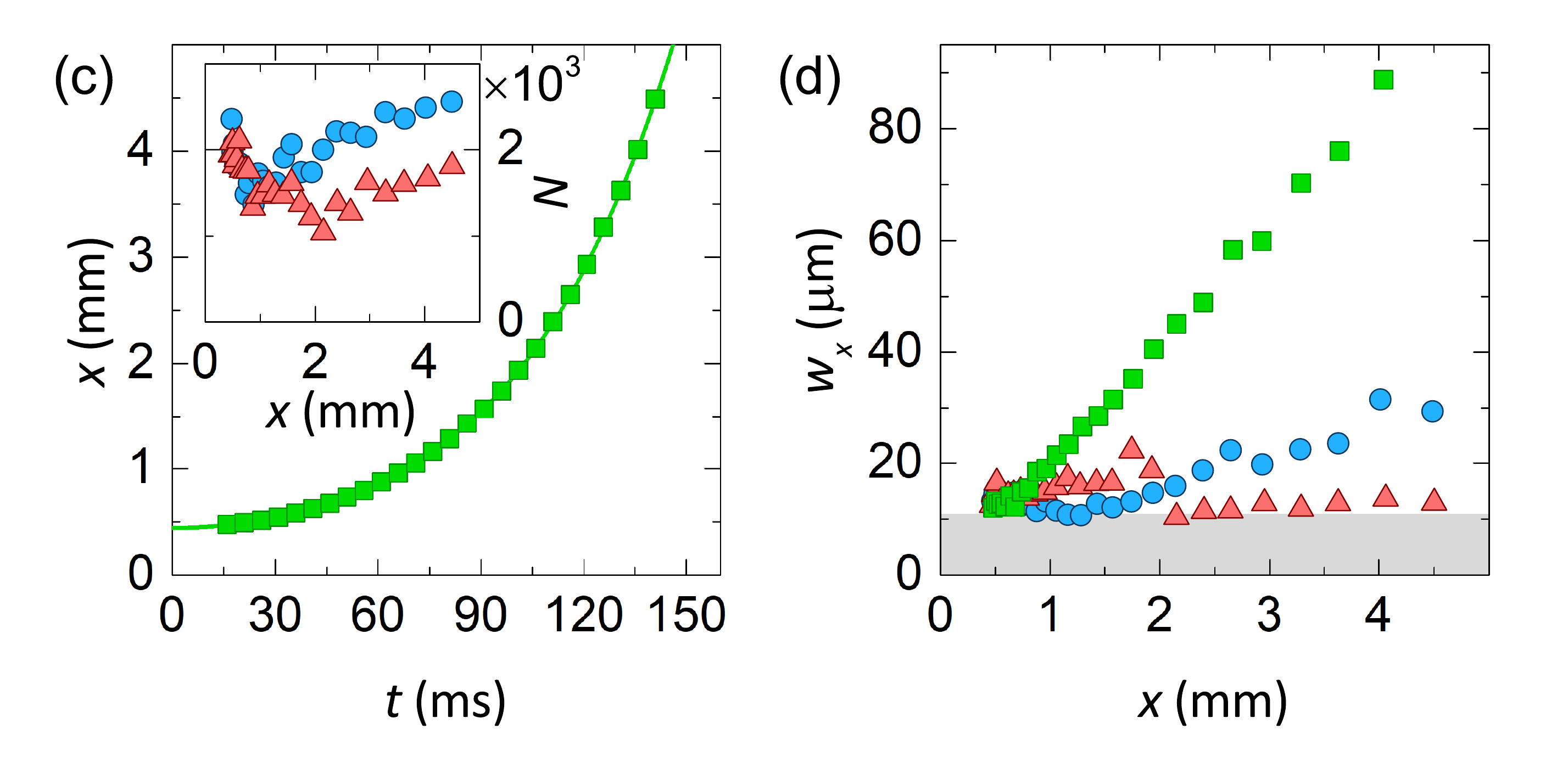}
\caption{\label{fig:2:Mag_field_comp} 
(a) Density absorption images of the BEC propagating in a channel for a slightly attractive interaction that doesn't compensate dispersion (initial scattering length $-6.7a_0$) and (b) solitonic interaction (initial scattering length $-9.9a_0$).
(c) Time dependence of the position of the BEC in the anti-trapping potential and 
(d) its width (Gaussian $\sigma$ radius) depending on position for insufficient attractive interaction (green squares, for the case in Fig.~(a)), solitonic interaction with (red triangles, the case in Fig.~(b)) and without (blue circles) the compensation of the magnetic field with position.
The gray area marks the widths below the resolution of our imaging system.
Inset in figure (c) shows how the fitted number of atoms changes with position for solitonic interaction with and without compensation.
}
\end{figure}

At this point the trap is tilted in the vertical direction because the magnetic field gradient was lowered during forced evaporation \cite{HungZhangGemelkeEtAl2008}. 
To symmetrize and strengthen the confinement for the soliton experiment we first adiabatically ramp the magnetic field gradient back to the full levitation value of 31.3~G/cm over 140~ms.
At this point the BEC is in a trap with frequencies $\{\omega_x,\omega_y,\omega_z\}=2\pi\cdot$\{40, 107, 114\}~Hz, with cesium atoms in the ground state.
We turn off one of the dimple beams, releasing the BEC into a quasi-one-dimensional channel along $x$ direction with radial frequency $\omega_r = \omega_y = 2\pi\cdot$107~Hz. 
In the axial direction of the channel there is a weak harmonic anti-trapping potential with "frequency" $2\pi\cdot 3.33$~Hz.
Cesium in the ground state $\lvert F=3, m_F=3 \rangle$ has a wide \textit{s}-wave Feshbach resonance with a zero crossing close to 17~G \cite{Chin2010}, which enables us to finely tune the interaction by changing the magnetic field (in App.~\ref{App:Interaction} we show how zero crossing at 17.26(20)~G is determined).
In the 5~ms before release into the quasi-1D channel we also ramp the magnetic field to a value $B_{sol}$, synchronizing the end of the ramp with the moment the BEC is released into the channel.
This way we go from a positive scattering length of 300$a_0$ to a negative scattering length $a$, which we vary from experiment to experiment.
Here $a_0$ denotes the Bohr radius and positive (negative) scattering lengths correspond to repulsive (attractive) interatomic interaction.
We then observe the evolution of the BEC in the quasi-1D channel using standard absorption imaging with 15~ms time-of-flight.

If the interaction between the atoms is repulsive or only slightly attractive, the BEC spreads in the axial direction of the channel (Figs.~\ref{fig:1:Soliton_interaction}(a) and \ref{fig:2:Mag_field_comp}(a)).
But if the interaction is just right, it exactly compensates dispersion and we get a soliton (Figs.~\ref{fig:1:Soliton_interaction}(b) and \ref{fig:2:Mag_field_comp}(b)).
If the interaction is even more attractive, the BEC usually separates into several solitons that form a so-called soliton train (Figs.~\ref{fig:1:Soliton_interaction}(c)-(e) and \ref{fig:3:SolTrain}).
Notice how in Fig.~\ref{fig:1:Soliton_interaction}(b)-(e) the individual solitons have fewer and fewer atoms as we go to stronger interaction. 
In a quasi-1D channel the highest number of atoms in a single soliton is given as $N_c = k_c a_r/|a|$, where $k_c\approx 0.67$ is a dimensionless interaction paramenter that depends on the axial and radial frequencies of the channel \cite{Parker2007}. 
$a_r = \sqrt{\hbar/m\omega_r}$ is the harmonic oscillator length of the channel with radial frequency $\omega_r$, mass of a single atom $m$ and the Plack constant $2\pi\hbar$ \cite{Donley2001, Parker2008}.
The number limit for a single soliton means that for higher interactions soliton trains with a higher number of solitons will form \cite{nguyen2017formation,Cornish2006}.
The reason we see only two solitons in Fig.~\ref{fig:1:Soliton_interaction}(d) and three solitons in Fig.~\ref{fig:1:Soliton_interaction}(e) are the collapses of the other solitons before the picture was taken.
Because of the repulsive interactions between solitons in soliton trains due to their relative phase, the trains with more solitons (more attractive interatomic interaction) spread out more along the axial direction of the channel, which can be seen by comparing Figs.~\ref{fig:1:Soliton_interaction}(c) and (e)
(further details about soliton train formation are provided in Sec.~\ref{Subsec:3:solTrains}).

To measure the dependence of the critical number of atoms on interaction, BECs with different atom numbers are created and formed into solitons inside a channel with radial frequency $\omega_r = 2\pi\cdot$101~Hz.
The number of atoms is measured by expanding a soliton with a strongly repulsive interaction and imaging an expanded soliton. 
These measurements are shown in Fig.~\ref{fig:1:Soliton_interaction}(f) for scattering lengths ranging from $-2 a_0$ to $-20 a_0$.
We can see that the critical number increases linearly with the inverse of scattering length as given by the equation above.

Because the stability of the solitons depends on the magnetic field, they can be used as a probe for magnetic field homogeneity in the experimental system.
A simple way to demonstrate this is by sending the soliton a long distance along the channel.
The weak anti-trapping potential in the axial direction of the channel, that is due to the magnetic field gradient used for levitation, has a form of a negative harmonic potential and can be written as $-\frac{1}{2}m\alpha^2 x^2$, where $\alpha=\sqrt{g^2 m/(3\mu_B B_0)}$ is its "frequency",
$x$ the coordinate in the axial direction of the channel, $g$ the gravitational acceleration, $\mu_B$ the Bohr magneton and $B_0$ the homogeneous magnetic field.
Putting a soliton in motion along the channel is therefore a matter of misalignment of its initial position along $x$ direction with respect to the center of the anti-trapping potential.
The center can be moved by changing the bias field in the $x$ direction with one of the compensation coils (see Appendix~\ref{App:ExperimentalApparatus}).

In this way we are able to observe the movement of the soliton over a distance of more than 4~mm.
It can reach a velocity of more than 100~mm/s before leaving the camera's field of view.
The trajectory of the movement is shown in Fig.~\ref{fig:2:Mag_field_comp}(c) and has a form $x(t) = x_0 \cosh (\alpha t)$, where $t$ is time.
By fitting position data with this function we determine the initial distance $x_0$ from the center of the quadrupole magnetic field and the "frequency" $\alpha=2\pi\cdot 3.33$~Hz which matches the calculated value for the used bias magnetic field $B_0\approx 17$~G.
Blue circles data in Fig.~\ref{fig:2:Mag_field_comp}(d) show the dependence of the BEC's width on its position. 
The BEC's width slightly increases, which is due to the inhomogeneity of the magnetic field on its path. 
Over the distance 4~mm the magnetic field increases by about 1.1~G, mostly due to the curvature of the field in horizontal direction, which accompanies the levitation gradient.
This equals the change in interaction from $-9.9a_0$ to $+53a_0$, ruining the soliton condition.
By changing the magnetic field dynamically as the BEC propagates along the channel we are able to compensate this spreading and maintain the soliton condition.
The red triangles in Fig.~\ref{fig:2:Mag_field_comp}(d) show the case where we change the magnetic field as a function that follows the motion of the BEC, always adjusting the scattering length. 
As we can see it works quite well for long distances and the BEC is truly a soliton in this case.
In the middle of the propagation in Fig.~\ref{fig:2:Mag_field_comp}(d) we see a sudden jump in width, which we suspect might be due to partial collapse of the soliton as in Ref.~\onlinecite{Donley2001} to adjust to the new, stronger interaction. 
This is also supported by the number of atoms data (inset in Fig.~\ref{fig:2:Mag_field_comp}(c)), where the red data shows fewer atoms than blue data from that point onward, indicating a partial collapse.
Its cause is the overcompensation of the magnetic field in the first part of propagation, which was decreased by 0.19~G instead of the magnitude of the inhomogeneity 0.15~G.
Interaction was therefore overcompensated to $-12.3a_0$ instead of $-9.9a_0$, which was the soliton condition in this case.

Due to limited resolution of our imaging system (seen as a band in Fig.~\ref{fig:2:Mag_field_comp}(d)) there is some uncertainty when determining the number of atoms in very narrow clouds.
This is evident on the inset in Fig.~\ref{fig:2:Mag_field_comp}(c), where the number of atoms appears to grow with propagation distance as the width increases.
Taking this into account, the number of atoms in solitons in Fig.~\ref{fig:1:Soliton_interaction}(f) was determined from an expanded soliton to ensure its accuracy.


\subsection{Soliton trains}
\label{Subsec:3:solTrains}

\begin{figure}[t]
\centering
\includegraphics[width=1\linewidth]{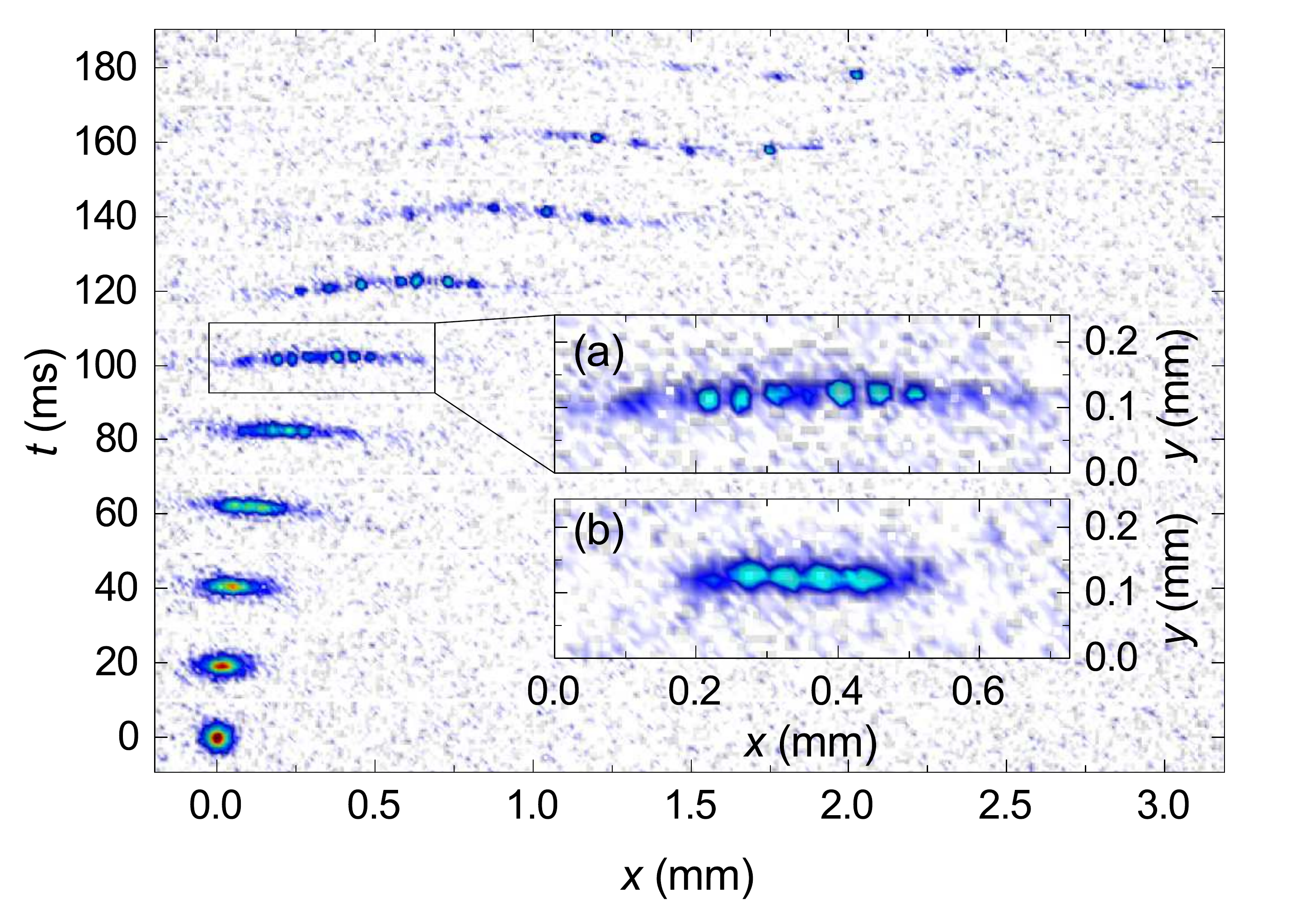}
\caption{\label{fig:3:SolTrain} 
Formation of a soliton train for scattering length $-13.5a_0$. 
We can see a soliton train gradually emerge from a slowly spreading BEC.
Inset (a) shows a soliton train created in a channel (enlarged picture of the train after 100~ms in the channel).
Inset (b) shows a fragmented BEC similar to a soliton train that was created outside the channel after 65~ms of time-of-flight.
}
\end{figure}

Besides single solitons, a train of multiple solitons can be produced from a single BEC \cite{strecker2002formation, Cornish2006, nguyen2017formation, Everitt2017}. 
The procedure is similar to the one described in the previous section, only the interaction $a$ is set to a stronger attractive value than for a single soliton.
A few examples of soliton trains are shown in Fig.~\ref{fig:1:Soliton_interaction}(c)-(e) and Fig.~\ref{fig:3:SolTrain}.
In previous experimental works soliton trains were usually created from an elongated BEC, but they can form from an almost circular BEC too.
Formation of a soliton train is shown in Fig.~\ref{fig:3:SolTrain}.
It happens due to modulation instability \cite{Zakharov2009} and has been explored both theoretically and experimentally \cite{Carr2004, nguyen2017formation, Everitt2017}.
Modulation instability exponentially enhances any random fluctuations in density, which leads to fragmentation of the BEC.
During the formation process the neighbouring solitons in a train acquire a relative phase $\pi$, which means they repel each other, inhibiting merger events and collapses, therefore increasing the stability of the train.

We start with a slightly elliptical BEC in the same trap as in the previous section, again releasing it into the channel along $x$ direction, with radial frequency $\omega_r=\omega_y=2\pi\cdot 107$~Hz.
In 5 ms before the release we ramp the scattering length from evaporation value $300a_0$ to the value $a\lesssim -8a_0$ for the formation of soliton trains.
In Fig.~\ref{fig:3:SolTrain} we see the evolution in the channel after release.
There is an additional time-of-flight where we release the atoms from the channel and let them evolve for 15 ms in free space before taking the picture.
Each picture is the result of a separate experimental run, due to the destructiveness of absorption imaging.
At first the BEC spreads out in the channel. 
Then, after 60~ms, we start seeing the beginnings of fragmentation.
In the next 20~ms the BEC separates into several solitons that can be clearly distinguished. 
These persist for about 20~ms, after which individual solitons start collapsing.
We found that in general smaller solitons were less stable to collapse than bigger ones.
The wavy shape of the soliton trains in pictures at longer times is due to slight initial misalignment of the atoms in the vertical direction which causes oscillations in the channel that are enhanced by the 15~ms of time-of-flight.

Interestingly, it is possible to produce a fragmented BEC similar to a soliton train without transverse confinement.
In another set of experiments we let the BEC evolve in the channel with repulsive interaction at scattering length $260a_0$ for 8~ms to elongate the BEC, after which we release it from the channel and switch the scattering length to $-68a_0$, i. e. a strongly attractive interaction.
After a 65~ms time-of-flight in this field we observe a fragmented BEC (Fig.~\ref{fig:3:SolTrain}(b)), despite the fact that the interaction was repulsive while the BEC was in the channel. 
This implies that geometric confinement is not needed for fragmentation as long as the BEC is already elongated into a quasi-1D shape.

Fragmentation of the BEC in free space can only occur for large negative scattering lengths. 
This is due to the two competing timescales involved.
The first one is the timescale for the formation of the soliton train $t_{train}=1/(2 n_{1D}|a| \omega_r)$, where $n_{1D}$ is the one-dimensional atom number density of the BEC confined in a quasi-1D channel with radial frequency $\omega_r$ \cite{nguyen2017formation}.
The other timescale is the spreading of a non-interacting BEC in the radial direction after being released from the quasi-1D channel $t_{spread}\approx 1/\omega_r$, which decreases the density of the BEC. 
The density decrease effectively decreases the interaction between atoms.
In free space the BEC can only fragment before the density decreases too much, which means that inequality $t_{train}\lesssim t_{spread}$ should hold.
For the case shown in Fig.~\ref{fig:3:SolTrain}(b) we estimate $t_{train}=0.7$~ms and $t_{spread}=1.5$~ms, confirming this scenario.
It is important to keep in mind, that the fragments of the BEC in free space are not solitons and are therefore not stable for long times.
Since there is no confinement to stabilize them, they dissipate after about 90~ms of TOF.

\subsection{Double soliton production and soliton collisions}
\label{Subsec:2:2xsol+collision}

\begin{figure}[t!]
\centering
\includegraphics[width=1\linewidth]{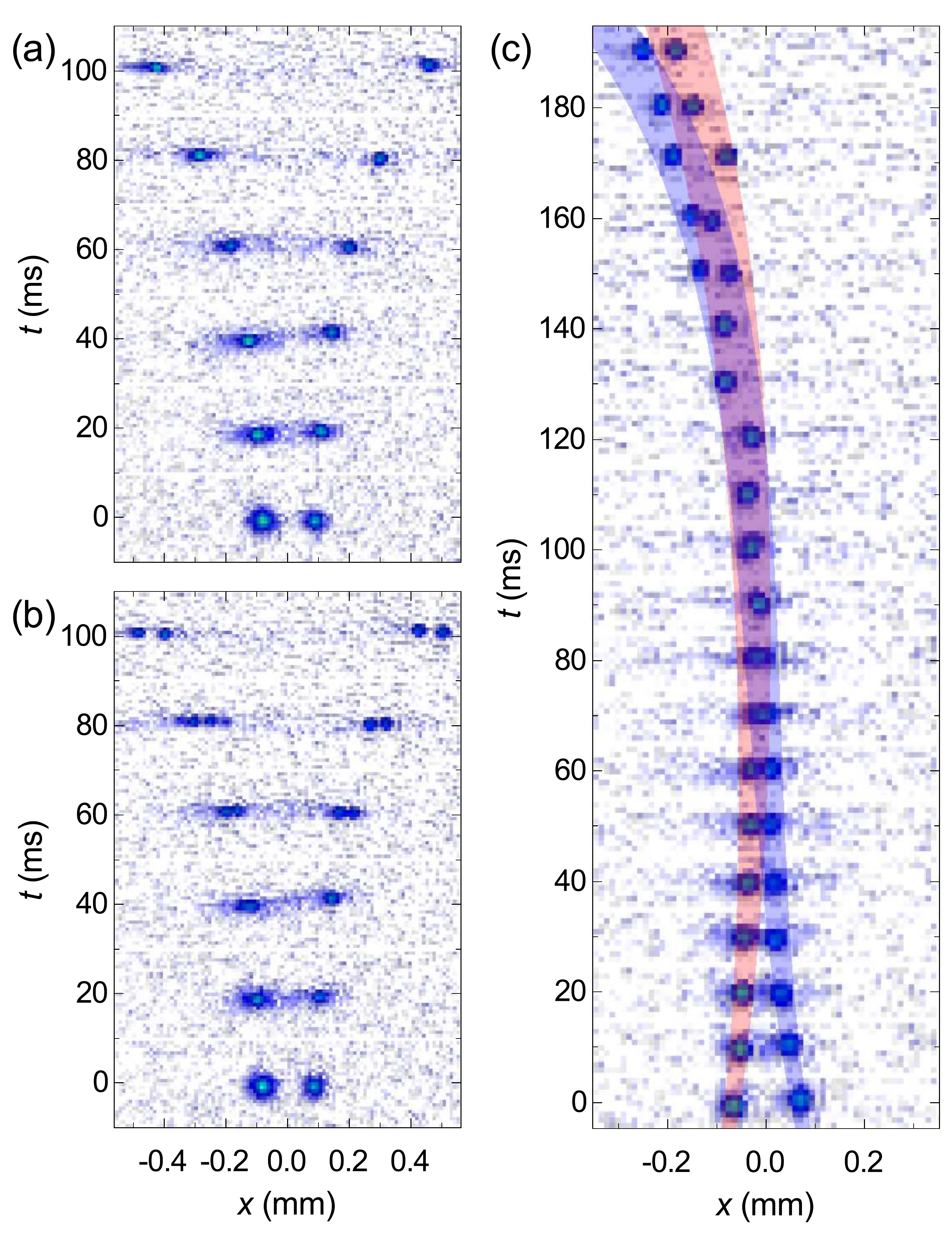}
\includegraphics[width=1\linewidth]{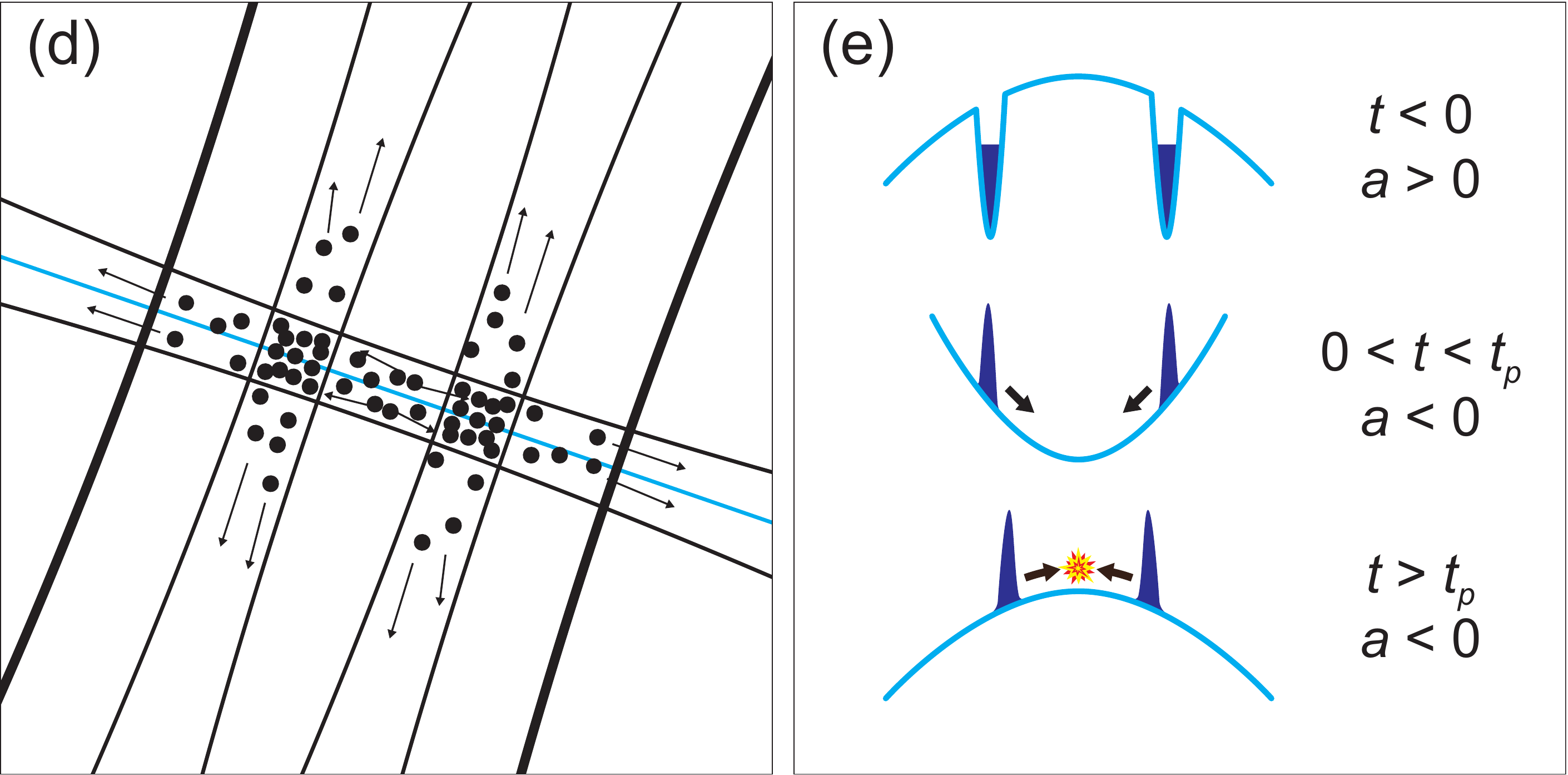}
\caption{\label{fig:4:SolCollision} 
(a) Propagation of two single solitons 
and (b) two soliton trains in the channel. 
(c) The slowest collision we observed ($<0.5$~mm/s). 
Two solitons pass through each other without changing. 
They move slightly to the left, indicating that the center of the potential is not exactly aligned with their center of mass. 
The red and blue belts mark possible trajectories for each soliton, taking into account initial velocity fluctuations.
(d) The geometry of the H trap configuration, with the large dipole beam (bold lines) used to collide solitons.
(e) Potential along the channel (blue line in Fig.~(d)) during three stages before the collision of solitons.
}
\end{figure}

Collisions between BEC solitons have been studied with simulations \cite{Khaykovich2006, Parker2008, Parker2009} and to some extent in the experiment \cite{nguyen2014collisions}.
In Ref.~\onlinecite{Parker2008} some interesting collisional events, such as mergers, collapse and population transfer between the solitons are simulated.
The paper also predicts the range of relative velocities and phases between the solitons where the collisions are stable, meaning the solitons pass through each other, and where they are unstable and one would see a collapse or a merger.
While most of these different collision events have already been demonstrated experimentaly in \cite{nguyen2014collisions}, the whole velocity-phase plane has yet to be explored.

To study collisions two BECs are needed.
One can split a single BEC into two by raising a repulsive dipole barrier in the middle  \cite{nguyen2014collisions}, in a magnetic double well \cite{Esteve2005}, using an optical grating  \cite{Kozuma1999, Kasevich1991, Cronin2009}, or by sending it through a narrow dipole potential \cite{Helm2014, Helm2015, Marchant2016}.
This way one can in principle control the relative phase between the two condensates' wavefunctions.
The other option is to create two BECs independently from the beginning.
This means creating two separate traps, performing evaporative cooling in both of them simultaneously, and thus producing two independent BECs with a random phase difference.

To produce two BECs we add another beam to our dimple trap, creating a letter H configuration (see Fig.~\ref{fig:4:SolCollision}(d), Appendix~\ref{App:PathToBEC}).
We follow the same procedure to get the BECs as before. 
For more efficient forced evaporation, it is helpful to keep the distance between the two parallel dimple beams as large as the reservoir dipole trap allows.
After evaporation we release the two BECs into the quasi-1D channel with radial frequency $\omega_r=2\pi\cdot 107$~Hz, change the interaction to attractive and watch them evolve.
By producing two BECs with a similar number of atoms we ensure the same interatomic interaction.
This makes the critical interaction for soliton formation for both BECs the same, enabling simultaneous creation of two single solitons as well as soliton trains (Figs.~\ref{fig:4:SolCollision}(a) and (b)).
If the atom numbers in the two BECs are too different, simultaneous solitons are impossible to achieve in a homogeneous magnetic field.

The two solitons are created symmetrically with respect to the center of the anti-trapping potential due to the magnetic field gradient mentioned above.
At first we were colliding them in the wide dipole trap beam that is perpendicular to the channel (bold beam in Fig.~\ref{fig:4:SolCollision}(d)). 
It creates a large harmonic potential in which the solitons can oscillate.
This way the collision velocity was always too high to observe any interesting collision events, such as mergers or collapses, because the atoms from different solitons need some time to interact with each other \cite{Parker2008}.
To decrease the velocity we give the solitons only an initial push of length $t_p$ with the dipole trap and then take advantage of the anti-trapping potential, letting them slow down as they climb up the potential towards the middle (Fig.~\ref{fig:4:SolCollision}(e)).
In this way the lowest achieved collision velocity was below 0.5~mm/s, which should be slow enough for mergers and collapses to happen according to Ref.~\onlinecite{Parker2008}.
Even though we observed hundreds of sufficiently slow collisions, we have yet to see a merger or collapse.
The slowest collision is shown in Fig.~\ref{fig:4:SolCollision}(c). 
The pictures are taken in the same way as described in the previous sections.
We see that for longer times the position of solitons fluctuates from picture to picture, which we attribute to an initial velocity caused by beam pointing fluctuations due to air flow along the beam path and subsequent refraction index fluctuations.
Encasing the beams completely should remedy this problem, making the experiments more repeatable.

Besides the collisional velocity one would have to control the relative phase between the solitons to experimentally measure phase diagrams from Ref.~\onlinecite{Parker2008}. 
We currently produce two solitons with random relative phase.
Precise control over the phase can only be achieved by carefully splitting one condensate into two, which is quite difficult to implement.
The alternative is to measure the relative phase after the collision with atom interferometry, but that only works for stable collision events.
Both can be achieved by introducing a narrow potential barrier perpendicular to the channel \cite{Helm2014}.

\section{Conclusions and outlook}
\label{Sec:4:Conclusions}

This paper presents our experimental setup, which has produced bright matter-wave solitons with $^{133}$Cs BEC.
We condense cesium atoms using techniques such as the dimple trick \cite{WeberHerbigMarkEtAl2003} and evaporation with magnetic gradient \cite{HungZhangGemelkeEtAl2008}, basing our experimental sequence on Ref.~\onlinecite{Groebner2016}.
Producing solitons with cesium atoms is relatively straightforward once one has a BEC because of a broad, low lying Feshbach resonance that can be reached with a single Helmholtz coil pair, allowing for precise control of the interatomic interaction.
We also show the production of soliton trains and dependence of the individual soliton's size on the interaction between atoms, which nicely follows the theoretical formula for the critical number of atoms in a soliton.
Astoundingly, we also observe a fragmentation of a BEC outside the channel, in free space. 
To our knowledge this has not been observed before and is an indication that modulation instability has a role in matter-wave physics beyond soliton formation in quasi-1D geometry.

Further, we have shown that we can produce two BECs at the same time and from them two completely independent solitons or soliton trains.
Improvement of the initial velocity uncertainty, implementation of phase control and optionally non-destructive imaging, would open up a possibility of an in-depth study of soliton collisions.

In addition, our setup allows for preparation of two clouds of different temperatures, which could be useful for studies of interaction between a BEC and thermal atoms of varying temperatures.
BEC could serve as a probe for the thermal atoms, possibly in different spin states, with different densities and number of atoms.

\begin{acknowledgments}

We would like to thank Francesca Ferlaino and Philipp Haslinger for their helpful comments.
We acknowledge the contributions to the experiment by Nejc Rosenstein, Gregor Bensa, Nejc Jan\v{s}a, Pavel Kos, Maj \v{S}kerjanc, Rok Venturini, Nina Sedej, Marion A. van Midden, Mark Ber\v{c}an, Matja\v{z} Gomil\v{s}ek, Ivan Kvasi\'{c} and Davorin Kotnik. 
This work was supported by the Slovenian Research Agency (research core fundings No. P1-0125 and No. P1-0099, and research project No. J2-8191).

\end{acknowledgments}


\appendix
\section{The Experimental Apparatus}
\label{App:ExperimentalApparatus}

\begin{figure}[t]
\includegraphics[width=1\linewidth]{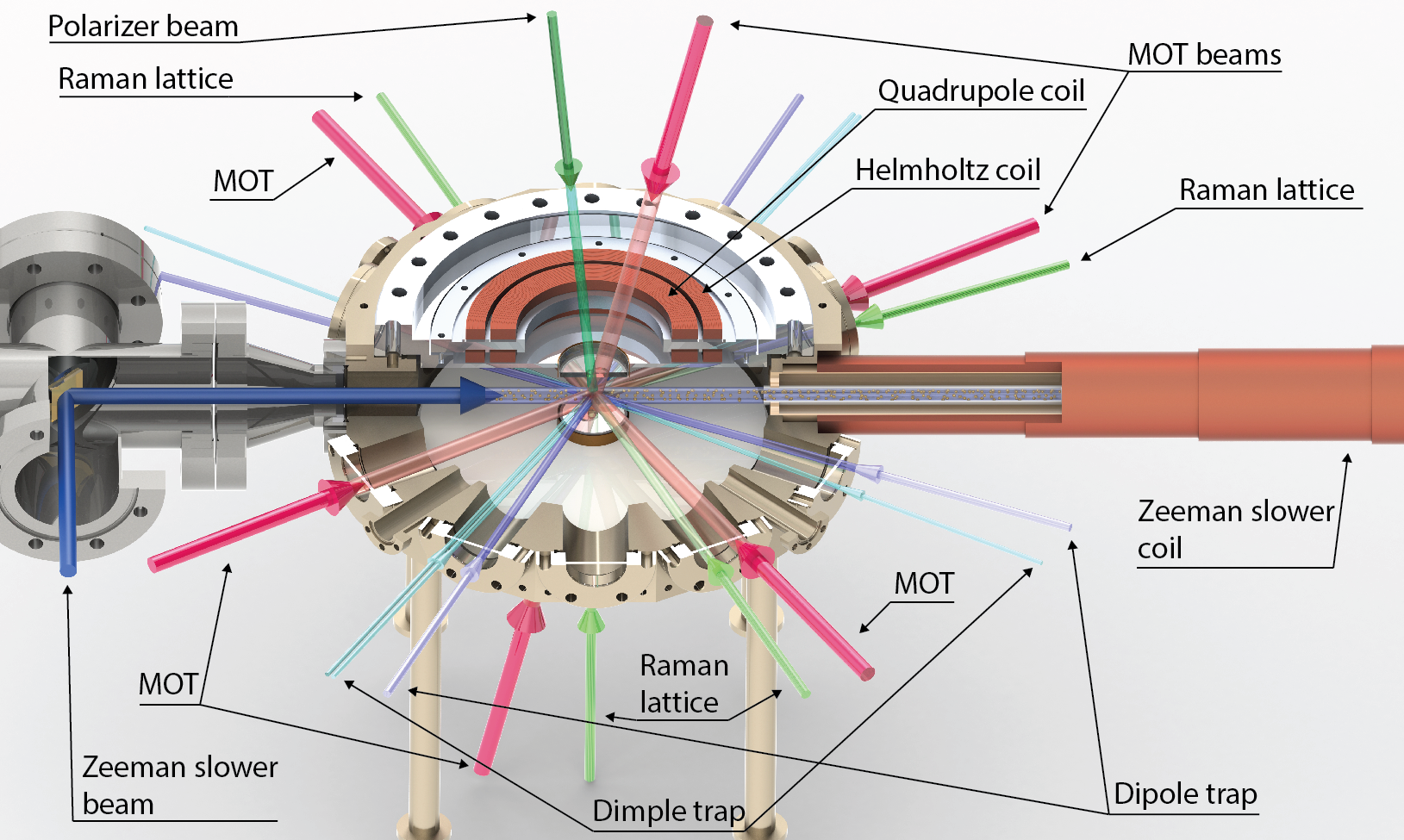}
\caption{\label{fig:A:vacuum_system} Main vacuum chamber and laser beams. 
}
\end{figure}

The centerpiece of the experimental setup is a stainless steel main vacuum chamber with the diameter of 31~cm (Fig.~\ref{fig:A:vacuum_system}). 
The laser beams used to cool, trap and image the atoms enter through 12 windows. 
A vacuum of $< 2\cdot 10^{-10}$~mbar is maintained by three ion pumps and a titanium sublimation pump.
Cesium atoms are emitted from dispenser rods made of Cs$_2$CrO$_4$, located in the oven that is connected to the main vacuum chamber via a narrow tube first 3~mm, then 5~mm and finally 10~mm in diameter that collimates the atomic beam.
A 70~cm Zeeman slower coil winds around this tube to create a magnetic field to compensate the Doppler shift caused by the slowing of the atoms.
Opposite of the atomic beam shines a Zeeman slower beam and its repumper. 
The atom flux can be shut off during experiments by a mechanical shutter in the form of a paddle that blocks the flow of atoms.
Two pairs of coils are wound around the top and bottom window of the main chamber, one in Helmholtz configuration and the other in anti-Helmholtz configuration. 
These produce homogeneous magnetic fields up to 250~G and magnetic field gradients up to 100~G/cm along the vertical direction, respectively.
To compensate the Earth's and any other stray homogeneous magnetic fields at the center of the main chamber, it is surrounded by three orthogonal pairs of compensation coils.
In the middle of the chamber six resonant beams cross to form a magneto-optical trap (MOT) together with the anti-Helmholtz coil.
In the same point four Raman lattice beams, the polarizer beam, the two dipole and dimple beams intersect.
When producing two BECs at he same time we have three dimple beams in H configuration (Fig.~\ref{fig:4:SolCollision}(d)). 
The dipole beams have $1/e^2$ radii of 670~$\mu$m, the dimple beam in $x$ direction 41~$\mu$m, and the two parallel dimple beams have radii 87~$\mu$m and 80~$\mu$m. 
To switch the beams on and off within microseconds we use acousto-optic modulators.
Main imaging is done with Andor iXon Ultra 888 EMCCD water cooled camera.
The imaging beam goes through the horizontal window perpendicular to the Zeeman slower and has a resolution of 10.9~$\mu$m. 
We also use an auxiliary imaging system in the vertical direction with 11~$\mu$m effective pixel size, using a CMOS camera IDS uEye ML, which mostly serves for beam alignment purposes.
Both imaging systems can serve for both absorption and fluorescence imaging.

The light for Zeeman slower, MOT, Raman lattice and imaging beams is derived from a Toptica TA PRO diode laser at 852 nm. 
The repumpers are all from a second 852 nm diode laser (Toptica DL 100).
For the large dipole trap we use 1070 nm 100 W IPG Nd:YAG fiber laser and for the dimple trap a 45 W 1064 nm Mephisto MOPA laser.
All beams are guided from the preparation table to the main chamber via optical fibers.

\section{Path to BEC}
\label{App:PathToBEC}

\begin{figure}[t!]
\centering
\includegraphics[width=1\linewidth]{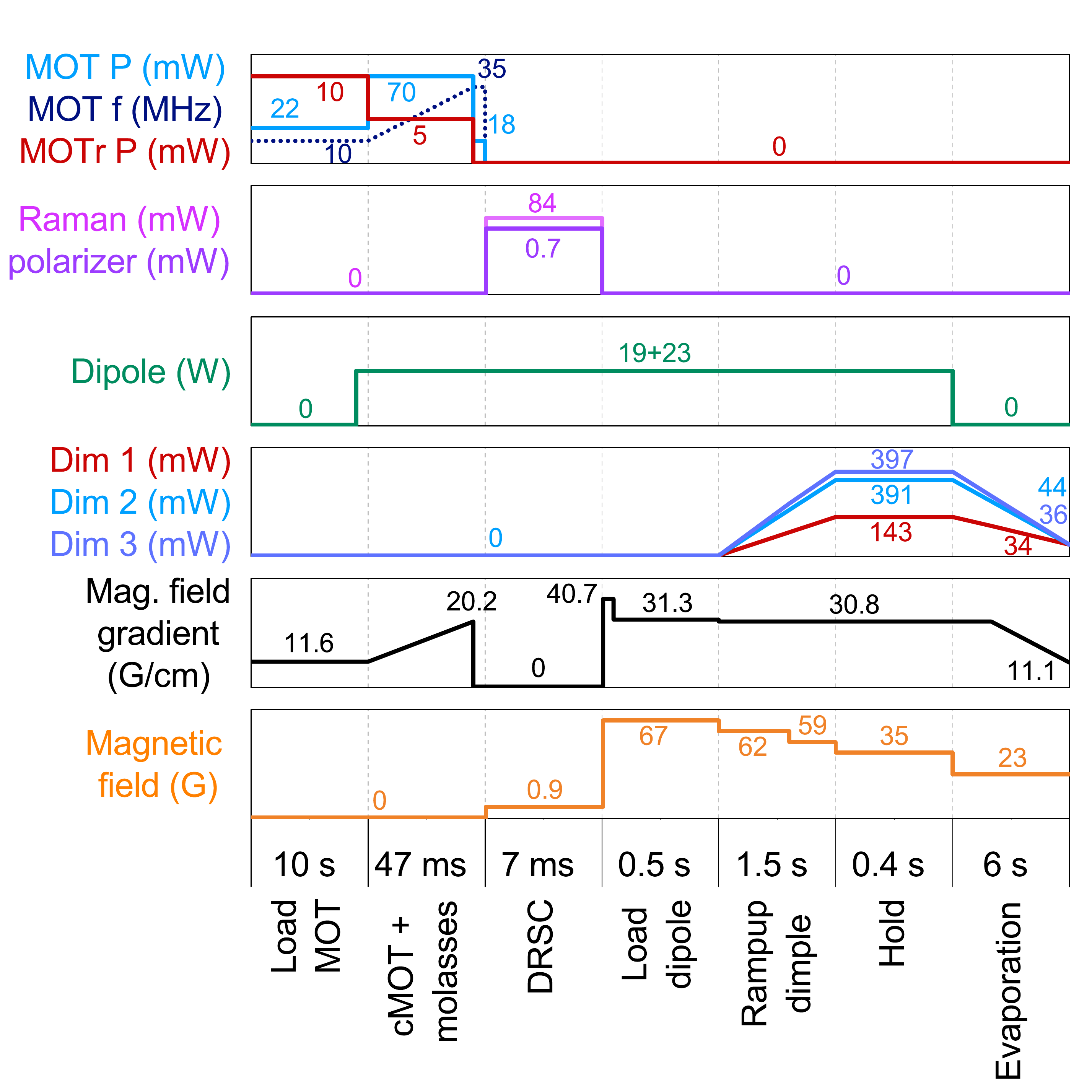}
\caption{\label{fig:B:sequence} Cooling sequence.
}
\end{figure}

The sequence for producing a BEC is shown in Fig.~\ref{fig:B:sequence}.
We start with atoms in the oven at 90~$^\circ$C that are collimated into a beam with velocities $\sim$ 200~$\mathrm{m/s}$. 
Opposite the atoms a beam with power 8~mW and diameter $\sim 1$~cm at the $F=4 \rightarrow F'=5$ transition with circular polarization slows the atoms down. 
Even though this is a closed transition, some atoms are still pumped into $F'=4$ excited state and spontaneously decay into $F=3$ ground state. 
Repumper with power 3~mW at the $F=3\rightarrow F'=4$ transition pumps these atoms back to $F=4$ to prevent them from getting lost.

As the atoms slow down, the frequency of the laser beams changes due to the Doppler shift.
This is compensated by the Zeeman shift provided by the Zeeman slower coil \cite{PhillipsMetcalf1982}.
When atoms exit the Zeeman slower and enter the main experimental chamber they are slow enough to be captured by a magneto-optical trap (MOT). 
The MOT is created by crossing three perpendicular pairs of beams with diameters $\sim 1.5$~cm and total power 22~mW. 
The incoming beams are $\sigma^+$ polarized and the retroreflected are $\sigma^-$ polarized. 
The magnetic field gradient used to load the trap is 11.6~$\mathrm{G/cm}$.
We usually load $\sim 60$ million atoms with temperature $\sim 70 \;\mathrm{\mu K}$ into the MOT over 10~s.

After loading we compress the atoms by ramping the magnetic field gradient to 20.2~$\mathrm{G/cm}$ and MOT beams' frequency detuning from $-10$~MHz to $-35$~MHz from the $F=4\rightarrow F'=5$ transition.
We also increase the power of MOT beams to 70~mW.
Before transferring them into the Raman lattice we turn off the gradient magnetic field, reduce MOT beams to a quarter of their former power and turn off the MOT repumper beams.
This creates an optical molasses that cools the atoms to $\sim 13\;\mathrm{\mu K}$ and transfers them to the internal ground state $F=3$ which is necessary for the degenerate Raman sideband cooling (DRSC) to work.
Because our vacuum chamber is made of stainless steel, inductance of the quadrupole and Helmholtz coils is large, which leads to switching times of about 6~ms.
This reduces the effectiveness of the MOT compression, because we have to wait several milliseconds for the gradient to reach zero, before we can trap the atoms in the Raman lattice.
During this time the atom cloud spreads, reducing the effect of compression.
An additional small booster coil would help change magnetic fields faster.

The Raman lattice is created by interfering 4 perpendicular beams along $z$, $\tilde{x}$ and $\pm \tilde{y}$ directions (21~mW each, resonant with $F=4\rightarrow F'=4$ transition), similar to the first realization in Ref.~\onlinecite{KermanVuleticChinEtAl2000} ($\tilde{x}$, $\tilde{y}$ lie at an angle $\sim 22^{\circ}$ to previously used $x$ and $y$, but still perpendicular to the vertical direction $z$).
The beams are linearly polarized in a way that they all have polarizations in the $\tilde{x}z$ plane.
There is also a circularly polarized polarizer beam, with power 0.7~mW and resonant with $F=3\rightarrow F'=2$ transition, slightly tilted with respect to the $z$ axis.
With Raman sideband cooling we can cool about 20 million atoms in the absolute ground state $|F=3,m_F=3\rangle$ to 1~$\mathrm{\mu K}$, which we then transfer into the large crossed dipole trap.

To improve phase-space matching for the Raman cooled atom cloud the dipole trap beams have large radii of 670~$\mathrm{\mu m}$ and powers 19.3~W and 22.6~W.
They are turned on for 0.5~s before the end of loading phase to prevent disturbing the atoms too much by turning them on suddenly at the end of the degenerate Raman sideband cooling (Fig.~\ref{fig:B:sequence}).
To nullify the velocity gain due to gravitational pull 
after turning off the Raman lattice beams, we first perform a 2~ms long upwards push in a magnetic field gradient 40.7~G/cm.
In order to stop the atoms from falling out of the dipole trap after that, they are levitated with a magnetic field gradient of 31.3~G/cm.
After 0.5~s of plain evaporation in the dipole trap we start ramping up the dimple trap. 
It is another crossed dipole trap (double crossed H trap for double BEC production), but with beams of much smaller radii (41~$\mu$m, 87~$\mu$m, 80~$\mu$m), located in the center of the large dipole trap (reservoir).
This changes the shape of the potential leading to a local increase of phase-space density inside the dimple trap \cite{Stamper-Kurn1998}.
The dimple beams are linearly ramped to their maximum powers (143~mW, 391~mW, 397~mW) in 1.5~s and held for another 400~ms to let the atoms thermalize. 
The magnetic field at the start of the ramp is lowered from 67~G to 62~G to reduce three-body losses and then after 900~ms to 59~G.
During the hold phase we set it to 35~G, because the increased density in the dimple trap would otherwise lead to a dramatic loss of atoms.
Then the large dipole trap is turned off and evaporation in the dimple trap begins.
At the beginning we have about $3\cdot 10^5$ atoms with phase-space density ($PSD$) of $5\cdot 10^{-3}$.
We evaporate in 23.3~G magnetic field, which corresponds to a scattering length of $300a_0$.
The powers of dimple beams are exponentially ramped down (to 34~mW, 44~mW and 36~mW) in 6~s and after 2~s we start with additional exponential rampdown of the magnetic field gradient to 11.1~G/cm (see Fig.~\ref{fig:B:sequence}).
The combination of the two techniques leads to efficient evaporation (average efficiency $\epsilon = \ln \frac{PSD}{PSD_0}/\ln \frac{N_0}{N}=2.4$, $PSD_0$ is initial phase-space density and $N_0$ initial atom number) and at the end of it we produce one or two BECs of 5000-10000 atoms depending on the exact evaporation parameters.
The evaporation efficiency is lower for the double BEC setup leading to $\sim20\%$ lower number of atoms in an individual BEC.


\section{Determining interatomic interaction}
\label{App:Interaction}


\begin{figure}[t!]
\centering
\includegraphics[width=1\linewidth]{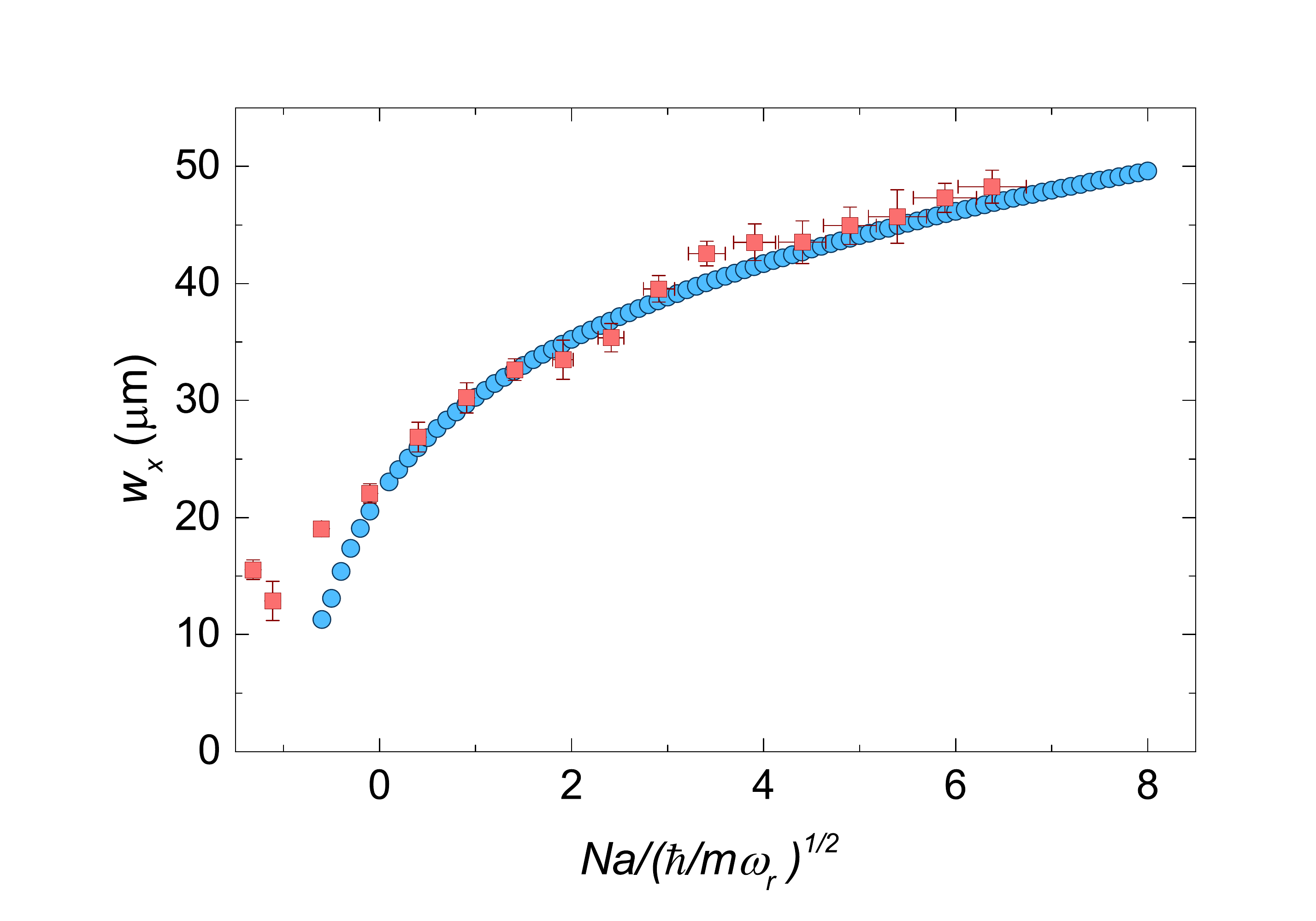}
\caption{\label{fig:C:simulated_data} Fit of measured width (red squares) to simulated width (blue circles) along the direction of the channel for different values of $k =Na/(\hbar/m\omega_r)^{1/2}$.
Each measured point is an average of 10 measurements.
}
\end{figure}

The system is described using the Gross-Pitaevskii equation (GPE) with a time-dependent potential term that describes the different stages of the experiment. 
The initial state is determined by evolution in the imaginary time in the presence of all trap potentials. 
The longitudinal trapping potential is then switched off, while the transverse trapping is maintained, to describe the evolution of the gas in the quasi-one-dimensional channel. 
In the model, we include the weak anti-trapping potential which has a component along the longitudinal direction of the system. 
We let the system evolve in real time for the duration $\tau$. 
Finally, we simulate the expansion of the system after all potentials are switched off (time-of-flight stage). 
From these we extract the width $\sigma_x = \sqrt{\langle x^2 \rangle}$ that we compare with the experimental values. 

This procedure is performed for a range of values for the nonlinearity parameter $k=Na/\sqrt{\hbar/m\omega_r}$ and for different evolution times $\tau$ to calibrate the scattering length $a$.
In calculations we use $L=\sqrt{\hbar/m\omega_r}$ as the length unit and $1/\omega_r$ as the time unit. 
The space is discretized on a 3D mesh. 
Most calculations have been performed in a volume $[-200,200] \times [-30,30] \times [-30,30]L^3$ on a mesh with $512 \times 64 \times 64$ points, with some additional calculations on a denser mesh (in particular for the case of strongly attractive interaction). 
The evolution in both imaginary and real time is based on a split-step method, with the kinetic energy terms evolved in reciprocal space and the potential and contact-interactions terms evolved in real space, using the FFTW library to perform the fast Fourier transforms. 
We used an algorithm with a second-order accuracy in the time step \cite{Suzuki1976,Yoshida1990} with $\delta t=0.001$ in most calculations.

We compare the simulation to measurement data and calibrate the zero of interaction by adjusting the constant $B_0$ in the s-wave Feshbach resonance $a(B)=a_{bg}\left( 1 - \frac{\Delta}{B-B_0} \right)$, where $a_{bg}$ is the background scattering length, $\Delta$ is the width of the Feshbach resonance, $B_0$ is the resonant magnetic field and $B$ is the magnetic field \cite{Chin2010}. 
After determining the field for zero interaction at 17.26(20)~G we can calculate interactions for other fields from the formula for the Feshbach resonance.
Measurement data fitted to simulated data is shown in Fig.~\ref{fig:C:simulated_data}.
To calibrate the magnetic field of the Helmholtz coils we used the narrow Feshbach resonances at 11.0~G, 14.4~G, 15.1~G, 19.9~G, 48.0~G, 53.5~G, 112.8~G and 131.1~G.

\bibliographystyle{apsrev4-1}
%

\end{document}